\documentclass[12pt,letterpaper]{article}
\usepackage{jheppub}

\usepackage{graphicx}
\usepackage{bbm}
\usepackage{amsmath}
\usepackage{amssymb}
\usepackage{mathtools}
\usepackage{amsthm}
\usepackage{array,multirow,esint,tabularx}
\usepackage[matrix,arrow,frame,import,curve,color]{xy}

\usepackage[labelformat=simple]{subcaption}

\usepackage{tikz}
\usetikzlibrary{calc}
\usetikzlibrary{arrows}
\usetikzlibrary{scopes}
\usetikzlibrary{intersections,through}
\usetikzlibrary{hobby}
\usetikzlibrary{decorations.pathmorphing,patterns}
\usetikzlibrary{math}
\usetikzlibrary{angles,quotes}
\usetikzlibrary{decorations.pathmorphing}
\usetikzlibrary{decorations.markings}

\makeatletter
\newcommand\@brcolwidth{0.67em}
\newenvironment{eqmatrix}{%
    \hskip-\arraycolsep
    \new@ifnextchar[\@brarray{\@brarray[\@brcolwidth]}%
}{%
    \endarray
    \hskip -\arraycolsep
}
\def\@brarray[#1]{\array{r*\c@MaxMatrixCols {M{#1}}}}
\makeatother

\newcolumntype{L}{>{$}l<{$}} 
\newcolumntype{C}{>{$}c<{$}} 
\newcolumntype{M}[1]{>{\hbox to #1\bgroup\hss$}l<{$\egroup}}

\newcommand{\normord}[1]{:\mathrel{#1}:}

\def\R{{\mathbb R}}
\def\Z{{\mathbb Z}}

\def\aoint{\varointctrclockwise}
\def\partiald{{\normalfont\textsf{\reflectbox{6}}}}
\def\CY{Calabi--Yau}

\def\be{\begin{equation}}
\def\ee{\end{equation}}

\DeclareFontFamily{U}{rsf}{}
\DeclareFontShape{U}{rsf}{m}{n}{
  <5> <6> rsfs5 <7> <8> <9> rsfs7 <10-> rsfs10}{}
\DeclareMathAlphabet\Scr{U}{rsf}{m}{n}

\def\ff#1#2{{\textstyle\frac{#1}{#2}}}

\newtheoremstyle{named}{0.75\baselineskip}{0.75\baselineskip}{\itshape}{}{\bfseries}{.}{.5em}{#3}
\theoremstyle{named}

\newcommand{\ket}[1]{\left|#1\right \rangle}
\newcommand{\bra}[1]{\left \langle #1\right |}

\newcommand{\eval}[3]{\left\langle #1 \left|#2\right|#3\right \rangle}

\begin{document}

\begin{titlepage}
\begin{flushright}
June 2024
\end{flushright}
\vspace{.5cm}
\begin{center}
\baselineskip=16pt
{\fontfamily{ptm}\selectfont\bfseries\huge
An Unreasonably Quick Introduction to String Theory,
  Conformal Field Theory and Geometry\\[20mm]}
{\bf\large Paul S.~Aspinwall
 } \\[7mm]

{\small

Departments of Mathematics and Physics,\\ 
  Box 90320, \\ Duke University,\\ 
 Durham, NC 27708-0320 \\ \vspace{6pt}

 }

\end{center}

\begin{center}
{\bf Abstract}
\end{center}

A very quick introduction to the bosonic string,
conformal field theory, the superstring and geometry. No
background in quantum field theory is assumed and the omissions are
vast. Based on four lectures at the 2024 Physical Mathematics of Quantum Field Theory
Summer School.  

\vfil\noindent

\end{titlepage}
\setcounter{tocdepth}{2}

\hrule
\tableofcontents

\bigskip\medskip
\hrule
\bigskip\bigskip

\section{Day 1}

Over the next 4 lectures we will rapidly whirl through the
construction of string theory and conformal field theory assuming as
little background as possible. Much of this is copied from the
standard codices
\cite{Polchinski:1998rq,Polchinski:1998rr,GSW:book,Gins:lect}. We
reference these once and for all here and direct you to these sources
for a far more complete derivation of what happens below.

\subsection{The Harmonic Oscillator}
We begin with the harmonic oscillator. This provides the foundational
concepts in field theory and strings, but we begin with the humble
single nonrelativistic particle.

\subsubsection{Classical}
The classical harmonic oscillator is viewed as a particle of mass $m$
with a displacement $x$.  The kinetic energy is given by
$\ff12m\dot x^2$, which we may also write as $p^2/2m$, where
$p=m\dot x$ is the momentum.

The definition of the harmonic oscillator is that the potential energy
is quadratic:
\begin{equation}
  V=\ff12m\omega^2x^2,
\end{equation}
and then the Hamiltonian $H$, i.e., total energy in this case, is the sum
of the kinetic and potential energies.

The classical solution of this is that $x(t)$ is a sine wave of
(angular) frequency $\omega$.

\subsubsection{Quantum}
In quantum mechanics we have a ``space of states'' given by a Hilbert
space $\Scr{H}$. The observables $x$ and $p$ are given by linear Hermitian,
or self-adjoint, operators acting on $\Scr{H}$. We then assert the
commutation relation 
\begin{equation}
  [x,p]=i\hbar.
\label{eq:xp}
\end{equation}

Dirac's trick involves introducing
\begin{equation}
  \begin{split}
    u&=\sqrt{\frac{m \omega}{2\hbar}}x+\frac{i}{\sqrt{2m\omega \hbar}}p\\
    u^\dagger&=\sqrt{\frac{m \omega}{2\hbar}}x-\frac{i}{\sqrt{2m\omega \hbar}}p.\\
\end{split}
  \label{eq:udef}
\end{equation}
Then
\begin{align}
	[u,u^\dagger]=1,\qquad 
	H=\left(u^\dagger u+\ff12\right)\hbar \omega.
\end{align}

If $\ket{E}\in \Scr{H}$ is an eigenvector of $H$, i.e.,
\begin{align}
	H\ket{E}=E\ket E,
\end{align}
then 
\begin{equation}
  \begin{split}
    Hu\ket{E}&=uH\ket{E}+[H,u]\ket E\\
    &=Eu\ket{E}+\hbar \omega [u^\dagger,u]u\ket E\\
	&=(E-\hbar \omega)u\ket E.
\end{split}
\end{equation}
So $u$ decreases $E$ by $\hbar \omega$. Similarly $u^\dagger$ increases $E$ by $\hbar \omega$.

Next, observe that
\begin{equation}
  \begin{split}
    \eval{E}{H}{E}&=E\left\|\ket{E}\right\|^2\\
    &=\bra{E}\left(u^\dagger u+\ff12\right)\ket E\hbar\omega\\
    &=\left (\|u \ket E\|^2+\ff12\left\|\ket{E}\right\|^2\right )\hbar\omega.
\end{split}
\end{equation}
Since our inner product is positive definite,
$H$ has eigenvalue $E\geq \frac 12\hbar \omega$, with equality if and
only if $u\ket E=0$.

So we can construct the Hilbert space $\Scr{H}$ as follows. Let a basis
element be the ground state $\ket g$ which is killed by $u$. Then the
rest of the basis is defined by $(u\dagger)^n\ket g$ for $n>0$
(perhaps suitably normalized). This is the ``Fock space'' version of a
Hilbert space.

Alternatively, the standard way to describe quantum mechanics is to say that
$\Scr{H}$ is the Hilbert space of square-integrable functions of
$x$. The operator $x$ is then simply multiplication by $x$ and the
momentum operator is
\begin{equation}
  p=-i\hbar \frac{d}{dx}.
\end{equation}
In this language one can show that the energy eigenstates of our
harmonic oscillator are given by
\begin{equation}
  \begin{split}
    \psi_n(x)&= \bra{x}(u^\dagger)^n)\ket g \\
	&=H_n\left (\sqrt{\frac{m\omega}{\hbar}}x\right )\exp\left(-\frac{m\omega}{2\hbar}x^2\right)\label{e.star},
\end{split}
\end{equation}
where $H_n$ are polynomials of degree $n$ in $x$ known as Hermite
polynomials. We'll need this in next lecture.

Let's set $\hbar$ to 1 from here on.

\subsection{A Quantum Violin String}

Now extend this to a vibrating violin string. 

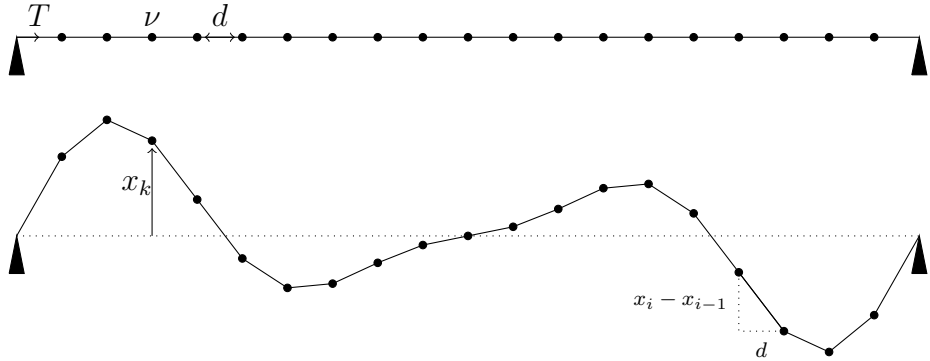
\begin{figure}
\begin{center}
\begin{tikzpicture}[scale=1]
\foreach \x in {1,...,19}
  \filldraw (0.6*\x,0) circle (0.05);
\draw (1.8,0) node[anchor=south] {$\nu$};
\draw [<->] (2.5,0) -- (2.9,0); 
\draw (2.7,0) node[anchor=south] {$d$};
\draw (0,0) -- (0.6*20,0);
\draw[->] (0,0) -- (0.3,0) node[anchor=south] {$T$};
\filldraw (-0.1,-0.5) -- (0,0) -- (0.1,-0.5) -- (-0.1,-0.5);
\filldraw (-0.1+20*0.6,-0.5) -- (0+20*0.6,0) -- (0.1+20*0.6,-0.5) -- 
   (-0.1+20*0.6,-0.5);
\end{tikzpicture}

\vspace{5mm}

\begin{tikzpicture}[scale=1]
\foreach \x in {0,...,20}
  \coordinate(p\x) at
    (0.6*\x,{0.2*sin(18*\x)+sin(2*18*\x)+0.5*sin(3*18*\x)});
\foreach \x [remember=\x as \lastx (initially 0)] in {1,...,19} {
  \draw (p\lastx) -- (p\x);
  \filldraw (p\x) circle (0.05);
}
\draw (p19) -- (p20);
\filldraw (-0.1,-0.5) -- (0,0) -- (0.1,-0.5) -- (-0.1,-0.5);
\filldraw (-0.1+20*0.6,-0.5) -- (0+20*0.6,0) -- (0.1+20*0.6,-0.5) -- 
   (-0.1+20*0.6,-0.5);
\draw[->] (p3 |- 0,0) -- ($(p3)-(0,0.1)$);
\draw ($(p3)-(0.2,0.6)$) node {$x_k$};
\draw[dotted] (0,0) -- (12,0);
\draw[dotted] (p16) -- (p16 |- p17) node[midway,anchor=east] 
{\scriptsize$x_i-x_{i-1}$} -- (p17)  node[midway,anchor=north] {\scriptsize$d$};
\draw (p16) -- (p17);
\end{tikzpicture}
\end{center}
\caption{A violin string made up of particles: at rest, and then plucked} \label{f.string}
\end{figure}

Consider the string of figure \ref{f.string}. Here, $\sigma$ is
horizontal displacement and $x$ is vertical displacement (which we
assume to be small in some sense). 
We'll consider the string to be made out of catgut molecules of mass
$\nu$ arranged with equal spacing $d$, and let $x_n$ denote the
displacement for the $n$th molecule.

Let the tension be $T$. We will take the limit of an
infinite number of such molecules. The potential energy between sites
is taken to be
\begin{align}
	U_i= \frac{T}{2d}(x_i-x_{i-1})^2.
\end{align}
Let's Fourier transform $x\rightarrow \tilde x$. The Hamiltonian can
then be shown to be
\begin{align}
	H=\sum_{n=1}^\infty \frac1{2\nu}\tilde
  p_n^2+\frac{2T}{d}\sum_{n=1}^\infty n^2\tilde x_n^2.
\end{align}
So we immediately recognize this as an infinite set of decoupled
harmonic oscillators, each with frequency proportional to $n$.  Let us
fix these various parameters so that the speed of sound along the
string is 1 and the frequency is $n$.

Accordingly we introduce an infinite set of pairs $u_n$ and $u_n^\dagger$ satisfying
\begin{align}
	[u_m^\dagger,u_n]=\delta_{m,n}.
\end{align}

It is standard in string theory to define
\begin{equation}
  \begin{split}
    \alpha_n&=\sqrt nu_n\qquad\text{for $n>0$}\\
	\alpha_{-n}&=\sqrt nu_n^\dagger \qquad\text{for $n>0$},\\
\end{split}
\end{equation}
so that we have a commutation relationship
\begin{equation}
  [\alpha_m,\alpha_n]=m\delta_{m+n}.
\end{equation}
Our notation is that a Kronecker delta with a single subscript is 1
precisely when that subscript is zero.  

\subsection{The Spectrum}

Let's analyze the spectrum. Denote the lowest energy state,
the ground state, as $\ket g$ again. So this is killed by
$\alpha_n$ for $n>0$:
\begin{align}
	\alpha_n\ket g&= 0 \qquad\text{if $n>0$}.
\end{align}
The Hamiltonian would appear to be
\begin{equation}
  H \stackrel{?}{=} \sum_{n=1}^\infty\left(\alpha_{-n}\alpha_n+\ff12n\right)
\end{equation}
which implies
\begin{align}
	H\ket g \stackrel{?}{=}\ff12\sum_{n=1}^\infty n\ket{g},
\end{align}
which is very unfortunate\footnote{Sometimes you might see
  claims by physicists at this point that $\sum n=-\ff1{12}$. We will not
  resort to such trickery here!}. Our so-called ground state has infinite
energy!

We will take the attitude that we don't actually care about the absolute
value of energy. What matters is {\em differences\/} in energies of
our states. Of course, this might not be reasonable in a theory with
gravity but let's ignore such issues. So let us reinterpret our
Hilbert space. Relabel the ground state to be $\ket0$ and
declare it to have zero energy. All other states, i.e., directions in
the Hilbert space, are obtained by applying raising operators
$\alpha_{-n}$, $n>0$, to $\ket0$. As such we build a 
Fock space again.  Our new Hamiltonian is then
\begin{align}
	H=\sum_{n=1}^\infty \alpha_{-n}\alpha_n.  \label{eq:Haa}
\end{align}

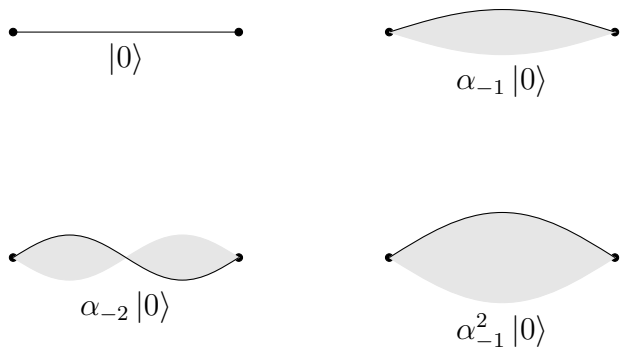
\begin{figure}
\begin{center}
  \begin{tikzpicture}
    \filldraw (0,0) circle (0.05);
    \filldraw (3,0) circle (0.05);
    \draw (0,0) -- (3,0) node[midway,anchor=north] {$\ket0$};
    \begin{scope}[xshift=5cm]
      \filldraw (0,0) circle (0.05);
      \filldraw (3,0) circle (0.05);
      \filldraw[gray!20] (0,0) sin (1.5,0.3) cos (3,0) sin(1.5,-0.3) cos(0,0);
      \draw (0,0) sin (1.5,0.3) cos (3,0);
      \draw (1.5,-0.3) node[anchor=north] {$\alpha_{-1}\ket0$};
    \end{scope}
    \begin{scope}[yshift=-3cm]
      \filldraw (0,0) circle (0.05);
      \filldraw (3,0) circle (0.05);
      \filldraw[gray!20] (0,0) sin (0.75,0.3) cos (1.5,0) sin(0.75,-0.3) cos(0,0);
      \filldraw[gray!20] (1.5,0) sin (2.25,0.3) cos (3,0) sin(2.25,-0.3) cos(1.5,0);
      \draw (0,0) sin (0.75,0.3) cos (1.5,0) sin(2.25,-0.3) cos (3,0);
      \draw (1.5,-0.3) node[anchor=north] {$\alpha_{-2}\ket0$};
    \end{scope}
    \begin{scope}[xshift=5cm,yshift=-3cm]
      \filldraw (0,0) circle (0.05);
      \filldraw (3,0) circle (0.05);
      \filldraw[gray!20] (0,0) sin (1.5,0.6) cos (3,0) sin(1.5,-0.6) cos(0,0);
      \draw (0,0) sin (1.5,0.6) cos (3,0);
      \draw (1.5,-0.6) node[anchor=north] {$\alpha_{-1}^2\ket0$};
    \end{scope}
  \end{tikzpicture}
\end{center}
\caption{Various string excitations.}
\label{f.stringexcitations}
\end{figure}

Our lowest level of excitation with energy 1 is then
$\alpha_{-1}\ket0$. Next we have energy 2 states given by
$\alpha_{-1}^2\ket 0$ or $\alpha_{-2}\ket 0$. These two states are different. See
Figure \ref{f.stringexcitations}. 

Our violin string has fixed boundary conditions at each end. Boundary
conditions in string theory lead to D-branes which are massively
important in almost all aspects of string theory. Sadly we will
completely ignore D-branes in these lectures.

\begin{figure}
\begin{center}
  \begin{tikzpicture}[scale=1.5]
    \draw (0,0) arc(0:85:1) .. controls (-0.98,1) .. (-1,1.05)
    .. controls (-1.02,1) .. (-1.087,0.996) arc (95:265:1)
    .. controls (-1.02,-1) .. (-1,-1.05) .. controls (-0.98,-1) .. (-0.913,-0.996)
    arc(-85:0:1);
    \draw[-stealth] (-0.8,1.2) arc(80:100:1.2)
    node[midway,anchor=south] {$\alpha_{-n}$};
    \draw[-stealth] (-0.8,-1.2) arc(280:260:1.2)
    node[midway,anchor=north] {$\tilde\alpha_{-n}$};
    \draw[-stealth] (-1,0) -- (-0.3,0) node[midway,anchor=south]
    {$\alpha_{0}$};
    \filldraw(-1,0) circle (0.04) node[anchor=east] {$\bar x$};
  \end{tikzpicture}
\end{center}
\caption{A closed string}
\label{f.closed}
\end{figure}
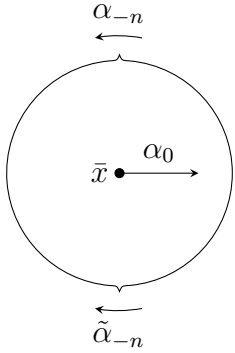

Instead, let's focus on closed strings. See Figure \ref{f.closed}. The
closed string differs from our violin string in two very important respects
\begin{enumerate}
\item The string is free to float about. So it has a center of mass
  position, $\bar x$, and momentum which are new degrees of freedom.
	\item The classical solutions are two decoupled sets of
          waves. One set moves counterclockwise around the string,
          while the other moves clockwise.
\end{enumerate}

For the violin string, our sine waves were functions of $\sigma$. So,
for the closed string we introduce
\begin{align}
	\sigma^\pm=t\pm \sigma,
\end{align}
and replace $\sigma$ with $\sigma^\pm$ for a doubled set of oscillators.
Actually, let's complexify things and use $-it$ instead of $t$, so that
\begin{align}
	\sigma^\pm=-it\pm \sigma.
\end{align}

We can now define the following new coordinates $z$ and $\bar z$:
\begin{align}
	z&=e^{i\sigma^+}\qquad \bar z=e^{i\sigma^+\-}.
\end{align}
As is common in complex analysis we will regard $z$ and $\bar z$ as
independent. Holomorphic functions of $z$ will represent the
counterclockwise oscillations on our closed string, while the
physically independent clockwise oscillations correspond to
antiholomorphic functions.

Quantization of our closed string therefore has 2 sets of raising and
lowering operators $\alpha_n$ and $\tilde\alpha_n$, as well as the
center of mass position $\bar x$ and momentum $p$. The commutation
relations are
\begin{equation}
  \begin{split}
  [\alpha_m,\alpha_n]&=m\delta_{m+n},\qquad
                       [\tilde\alpha_m,\tilde\alpha_n]=m\delta_{m+n}\\
    [\tilde\alpha_m,\alpha_n]&=0\\
    [\bar x,p] &= i.
    \end{split}  \label{eq:alphacom}
\end{equation}
A little algebra, going back to our definition (\ref{eq:udef}), then yields
\begin{align}
	x(z,\bar z)=\bar x+\frac{p}{4\pi i T}\log
  (z\bar z)+\frac{i}{2\sqrt{\pi T}}\sum_{n\neq 0}\frac 1n(\alpha_n
  z^{-n}+\tilde \alpha_n\bar z^{-n}).
  \label{eq:x}
\end{align}

Note we have yet to define $\alpha_0$. It is very convenient to set
\begin{equation}
  \alpha_0 = \tilde\alpha_0 = p,
\end{equation}
and $T=\frac{1}{4\pi}$. One then has simple relationships
\begin{align}
  \begin{split}
	i\partial x&=\sum_{n\in \mathbb Z}\frac{\alpha_n}{z^{n+1}}\\
	i\bar\partial x&=\sum_{n\in \mathbb
                         Z}\frac{\tilde\alpha_n}{\bar z^{n+1}},
\end{split} \label{eq:alphasum}
\end{align}
where $\partial$ denotes $\partial/\partial z$ and $\bar\partial$
denotes $\partial/\partial\bar  z$.

Let us define our ground state to satisfy $\alpha_0\ket 0=0$ so that
it corresponds to zero momentum as well as no oscillations. (For a full
description we also need to add zero oscillator states $\ket{0;p}$
with nonzero momentum $p$.)

What we have here is a perfectly good (nonrelativistic) quantum
mechanical description of a vibrating string. So what's all the fuss
about string theory then if we can describe it with so little work?
Well, this string is not really the one we think of when we say
``string theory''. 

\subsection{Fundamental Strings}

A truly {\em fundamental\/} string is not made of cat
molecules\footnote{Perhaps we should emphasize that ``catgut'' is also not
  made out of cats.} or any other type of particle. It should be
purely identified as itself as a string without any
constituents. Quantum mechanically this makes a huge difference.

The way to state this is that $\sigma$ (or $\sigma^\pm$)
reparametrizations are a ``gauge symmetry''. A gauge symmetry is a
symmetry of the description of our model, but one that nature is
unaware of. Our model makes explicit use of $\sigma$, indeed we don't know
how to describe it without such a parametrization, but a fundamental
string must have no kind of dependence on this parameter. 

Reparametrizations of the circle are invertible maps $S^1\to S^1$
which we will assume are differentiable. These form the group
$\text{Diff}(S^1)$.

We write a reparametrization as
\begin{align}
	\sigma \to \sigma'=\sigma'(\sigma).
\end{align}
Classically, if we think of our universe as having a single spacial
dimension with coordinate $\sigma$, then a ``field'' is simply a
function of $\sigma$.  If $\varphi$ is a ``scalar field'' under a
reparametrization we would replace it by a field $\varphi'$ such that
\begin{align}
	\varphi'(\sigma')=\varphi(\sigma),
\end{align}
For the equivalent of vector fields, etc., we consider $A$ such that
\begin{align}
  A'(\sigma')=\left(\frac{d\sigma}{d\sigma'}\right)^hA(\sigma)
  \label{eq:prim}
\end{align}
where $h$ is called the ``weight". Such fields $A$ are called ``primary
fields''.

The Lie algebra of a group corresponds to elements infinitesimally
close to the identity. So consider 
\begin{equation}
  \sigma \rightarrow \sigma + i\varepsilon e^{in\sigma}.
\end{equation}
for small $\varepsilon$ and only consider first order effects in
$\varepsilon$. This induces
\begin{equation}
  A(\sigma')=A(\sigma)+i\varepsilon e^{in\sigma}\frac{d A}{d\sigma}.
\end{equation}

Thus the generators of our Lie algebra can be written as
\begin{equation}
  L_n=i e^{i n \sigma}\frac{d}{d\sigma}
\end{equation}
and it is easy to see these obey the commutation relation, i.e., Lie bracket:
\begin{equation}
  [L_m,L_n]=(m-n)L_{m+n}.
\end{equation}
This Lie algebra is called the ``Witt Algebra''.

Now, for a change induced by $L_m$,
\begin{equation}
\begin{split}
\delta A &= A'(\sigma) - A(\sigma)\\
  &= A'(\sigma'-i\epsilon e^{im\sigma}) - A(\sigma)\\
  &= A'(\sigma') -i\epsilon e^{im\sigma}\frac{dA}{d\sigma} -
  A(\sigma)\\
  &= \left(\frac{d\sigma}{d\sigma'}\right)^h A(\sigma) -i
    \epsilon e^{im\sigma}\frac{dA}{d\sigma} -
  A(\sigma)\\
  &= \left(1-\epsilon me^{im\sigma}\right)^{-h}A(\sigma)  -i
    \epsilon e^{im\sigma}\frac{dA}{d\sigma} -
  A(\sigma)\\
  &= \epsilon e^{im\sigma} \left(-i\frac{d}{d\sigma} + mh\right) A.
\end{split}
\end{equation}
So, if we take a Fourier decomposition of our primary field
\begin{align}
	A(\sigma)=\sum_n A_ne^{-in\sigma}
\end{align}
then
\begin{equation}
\begin{split}
 \epsilon^{-1}\delta A &= \sum_n e^{im\sigma}  \left(-i\frac{d}{d\sigma} +
     mh\right) A_ne^{-in\sigma}\\
&= \sum_n (-n +mh)A_ne^{i(m-n)\sigma}\\
&= \sum_n (m(h-1) -n) A_{m+n}e^{-in\sigma}
\end{split}  \label{eq:delta-A}
\end{equation}
This implies, if we think of $A$ as an operator,
\begin{align}
	[L_m,A_n]&=(m(h-1)-n)A_{m+n}.  \label{eq:prim-com}
\end{align}
In particular,
\begin{equation}
  [L_0,A_n] = -n A_n.
\end{equation}
We view this as a generalization of ``raising and lowering'' operator
statements. If $L_0$ is viewed as a kind of Hamiltonian, then $A_n$'s
with positive subscripts are the lowering operators, while negative
subscripts produce excitations.

Rather than a single $\sigma$, for a our closed string we have
$\sigma^\pm$ which we have written in terms of a complex coordinate
$z$. So we really have two reparametrization groups and two Lie
algebras with generators
\begin{equation}
  L_n= -z^{n+1}\partial,\quad\tilde L_n=
  -\bar z^{n+1}\bar\partial.
\end{equation}
Written this way, the $L_n$'s form the Lie algebra of
(meromorphic) complex reparametrizations. Such mappings of the complex
plane are {\em conformal}, i.e., angle preserving. 

So, by starting out trying to analyze string theory, we've ended up
with a model on the complex plane where we are interested in invariance
under conformal maps. This is how ``conformal field theory'' is part
of string theory.

We want to think of how the $L_n$'s (and $\tilde L_n$'s) act on our
quantum mechanical string. We've described our string in terms of a
Fock space $\Scr{H}$ in which we rather delicately removed an infinity
by suitably defining our ground state. It would be nice if we could
describe the action of the Witt algebra purely in terms of this
language so that we don't disturb the validity of this construction.

In other words, can we write the $L_n$'s purely in terms of the only
operators we have, namely the $\alpha_n$'s? First note that if $A$ is a
primary field of weight $h$ then the change of coordinates from
$\sigma^+$ to $z$ means that the Fourier series becomes (up to
factors we can drop)
\begin{equation}
  A(z) = \sum_n\frac{A_n}{z^{n+h}}.
\end{equation}
Given how derivatives transform, we see from (\ref{eq:prim}) that
$i\partial x(z)$ has $h=1$ assuming it is a primary field, and that its
Fourier modes are therefore $\alpha_n$. 
Thus, (\ref{eq:prim-com}) gives
\begin{align}
	[L_m,\alpha_n]=-n\alpha_{m+n}.  \label{eq:Lalpha}
\end{align}

So our goal is to build $L_m$ from the $\alpha_n$'s such that this is
true. Now, a bracket is a derivation, i.e., it obeys Leibniz rule, and
we have (\ref{eq:alphacom}). This implies $L_m$ must be quadratic in
the $\alpha$'s. Indeed the following attempt yields (\ref{eq:Lalpha})
\begin{align}
	L_m\stackrel{?}{=}\ff12\sum_{k\in \mathbb Z}\alpha_{m+k}\alpha_{-k}
\end{align}
But then we would have
\begin{align}
  L_0\ket{0}\stackrel{?}{=}\ff12\sum_{k\in \mathbb Z}
  \alpha_k\alpha_{-k}\ket 0\stackrel{?}{=}\ff12\sum_{k=1}^\infty k\ket 0.
\end{align}
Our old divergent nemesis has returned. But actually we are being a
bit too quick here. Defining $L_n$ purely by its commutator with
something means we could always have added an arbitrary constant
term. Furthermore, changing the order of the $\alpha_n$'s in the the
definition of $L_m$ would shift $L_0$ by a constant. So we are
completely free to reorder and this can remove the divergence. Define
``normal ordering'':
\begin{equation}
\normord{\alpha_m\alpha_n}=\begin{cases}
		\alpha_m\alpha_n &\text{ if $m\leq n$}\\
		\alpha_n\alpha_m &\text{ if $m>n$}
	\end{cases}
\end{equation}
and then define
\begin{align}
	L_m=\ff12\sum_{k\in \mathbb
  Z}\normord{\alpha_{m+k}\alpha_{-k}}.  \label{eq:Lboson}
\end{align}
This does not mess up (\ref{eq:Lalpha}) but now
\begin{equation}
  L_n\ket 0 = 0, \quad \forall n\geq -1,
\end{equation}
and $L_n\ket0$ with $n<-1$ only has a finite number of excited
oscillator modes, so our divergences have gone away. Also, $L_0$
coincides with the Hamiltonian (\ref{eq:Haa}).

Now, however, with sufficient stamina and care, one can show this
normal ordering has changed the algebra to
\begin{align}
	[L_m,L_n]=(m-n)L_{m+n}+\ff{1}{12}m(m^2-1)\delta_{m+n}.
\end{align}
(We encourage this as a homework exercise.)

Note we still have freedom to add an arbitrary {\em finite\/} constant to
$L_0$. This comes back to haunt us shortly.

\subsection{The Virasoro Algebra}
The Virasoro Algebra is defined as having a basis $L_n$, $n\in\Z$
together with $c$ and brackets
\begin{align}
	[L_m,L_n]=(m-n)L_{m+n}+\frac{c}{12}m(m^2-1)\delta_{m+n},
\end{align}
and any bracket with $c$ equal to zero. That is, $c$ is a {\em
  central\/} element. The Virasoro algebra is thus a {\em central extension\/}
of the Witt algebra.

In any representation of this algebra, it is common to take the
generator $c$ to be a multiple of the identity operator. Then,
reinterpreting notation a little, we think of $c$ as this multiple. So
$c$ is then a number. This latter convention is standard in the
physics literature and then $c$ is called the {\em central charge\/} of
the Virasoro algebra.

So we have shown that the reparametrization of our quantum mechanical
string corresponds to a Virasoro algebra with central charge $c=1$.

Now, to be more general we could think of a string vibrating in $D$
dimensions by indexing our displacement as $x^\mu$,
$\mu=1,\ldots,D$. We would then have oscillators $\alpha_n^\mu$
obeying commutation relations
\begin{equation}
  [\alpha_m^\mu,\alpha_n^\nu] = m\delta^{\mu,\nu}\delta_{m+n}.
\end{equation}
It is then an easy matter to show that the resulting model has central
charge
\begin{equation}
  c = D.
\end{equation}
Indeed, the central charge is additive in the sense that if we take
the combination of two (non-mutually-interacting) quantum mechanical
systems with a Virasoro algebra then the combined system has a central
charge equal to to sum of its parts. As such, the central charge is a measure
of the ``content'' of a system.

So we think of our string theory having a D-dimensional ``target
space''. But is this space or spacetime? We really have been cheating
somewhat so far compared to a more careful analysis of string
theory. We have been able to get away with a non-relativistic view of
string oscillations and we have not really addressed time. The correct
treatment is to view a two dimensional ``worldsheet'' $\Sigma$ with
local parameters $\sigma$ and $\tau$ and map this into spacetime with
coordinates $x^\mu$, one of which is timelike. So we have a kind of
time coordinate on the worldsheet and another time coordinate in
spacetime. Furthermore we now have to worry about the bigger group of
reparametrizations of the two-dimensional string
worldsheet. Fortunately none of all this makes too much of a
difference to our simplified presentation. We refer to a standard
treatment such as \cite{Polchinski:1998rq} for the full story. What
matters is that the interesting part of the reparametrization symmetry
is still the (left and right-moving) Virasoro algebra.

Now we understand the reparametrization symmetry of the initial model, we
need to treat it as a gauge symmetry and get rid of it. A na\"\i ve
approach would be to look at the part of the Hilbert space invariant
under the Virasoro algebra. Sadly this produces a completely empty
theory. Instead we should play the game of living correctly in a Fock
space picture of the Hilbert space. The nicest way of doing this is
via BRST quantization which, sadly, we have nowhere near enough time
to cover.

What BRST quantization does is to add more content to the theory in
the form of ``ghosts''. These form their own representation of the
Virasoro algebra but this time with $c=-26$. It is very satisfying
that this central charge is negative as gauging a symmetry should be
taking away degrees of freedom in some sense. Finally one shows that
the content of the resulting theory is only nontrivial if the central
charges of the ghosts and spacetime parts cancel and so we end up with
the result that we require
\begin{equation}
  D=26.
\end{equation}
That is, this particular model of a fundamental string only works if
spacetime has 26 dimensions.

Remember that finite constant we could add to $L_0$? It turns out that
the BRST process fixes this constant too, but in such a way that the ground
state $\ket0$ has negative energy squared. This is physically really
bad and can be interpreted as predicting ``tachyons''. Fortunately
the superstring that we see on day 3 fixes this. 

\section{Day 2}
\subsection{Path Integrals}

So far we have a non-interacting string. To get something more
interesting we need some kind of field theory picture. In particular,
we will describe two-dimensional conformal field theory, which is, in
many ways, the simplest of all quantum field theories to analyze. 

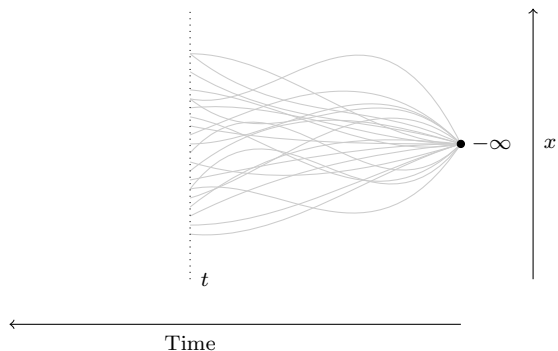
\begin{figure}
\begin{center}
\begin{tikzpicture}[scale=1.2]
\begin{scope}[color=black!20]
\draw (0,1) .. controls (1,1) and (2,0) .. (3,0);
\draw (0,1) .. controls (1,0.2) and (2,0.5) .. (3,0);
\draw (0,0.8) .. controls (1,0.2) and (2,0.3) .. (3,0);
\draw (0,0.6) .. controls (1,0.5) and (2,-0.2) .. (3,0);
\draw (0,0.5) .. controls (1,-0.3) and (2,1) .. (3,0);
\draw (0,0.5) .. controls (1,0.3) and (2,2) .. (3,0);
\draw (0,0.4) .. controls (1,0.5) and (2,0) .. (3,0);
\draw (0,0.3) .. controls (1,0.1) and (2,-1) .. (3,0);
\draw (0,0.2) .. controls (1,0) and (2,0) .. (3,0);
\draw (0,0.1) .. controls (1,0.5) and (2,1) .. (3,0);
\draw (0,0) .. controls (1,0) and (2,1) .. (3,0);
\draw (0,-0.1) .. controls (1,1) and (2,-1) .. (3,0);
\draw (0,-0.2) .. controls (1,-0.5) and (2,-0.3) .. (3,0);
\draw (0,-0.3) .. controls (1,0.5) and (2,0.3) .. (3,0);
\draw (0,-0.4) .. controls (1,-0.2) and (2,0) .. (3,0);
\draw (0,-0.5) .. controls (1,-0.3) and (2,-1.5) .. (3,0);
\draw (0,-0.5) .. controls (1,0.8) and (2,-1.3) .. (3,0);
\draw (0,-0.6) .. controls (1,-0.3) and (2,0.8) .. (3,0);
\draw (0,-0.7) .. controls (1,0.1) and (2,0.1) .. (3,0);
\draw (0,-0.8) .. controls (1,-0.3) and (2,-0.1) .. (3,0);
\draw (0,-0.9) .. controls (1,-0.9) and (2,-0.3) .. (3,0);
\draw (0,-1) .. controls (1,-1.1) and (2,-0.3) .. (3,0);
\end{scope}
\filldraw (3,0) circle (0.04) node[anchor=west] {\scriptsize $-\infty$};
\draw[dotted] (0,-1.5) node[anchor=west] {\scriptsize $t$} -- (0,1.5);
\draw[->] (3,-2) -- (-2,-2);
\draw (0,-2) node[anchor=north] {\scriptsize Time};
\draw[->] (3.8,-1.5) -- (3.8,1.5);
\draw (3.8,0) node[anchor=west] {\scriptsize $x$};
\end{tikzpicture}
\end{center}
\caption{A Wavefunction in a Path Integral.} 
\label{f.path integral}
\end{figure}

We begin with Feynman's interpretation of quantum mechanics via path
integrals \cite{FH:path}. For a quantum mechanical particle living on a real line
$\R$ parametrized by $x$ we have a {\em wavefunction\/}
$\psi(x)$. $\psi(x)$ represents an {\em amplitude\/} while
$|\psi(x)|^2$ represents the {\em probability\/} density of a particle
being at $x$.

In the path integral approach we assume we have some kind of boundary
condition lurking in the infinite past. Fixing a time $t$, we can then
compute the wave function at this time by an integral
\begin{align}
	\psi_t(x') =\int_{-\infty}^{x',t}\Scr{D}x\,e^{iS},
\end{align}
where we integrate over all possible paths starting in the infinite
past and ending at our point of interest as shown in
figure~\ref{f.path integral}. Note we have time flowing from right to
left in this picture, which we need to be consistent with the usual bra-ket
notation. The weighting factor is associated to the {\em action\/} $S$
of a given path. Analytically this path integral is awkward to define
and we use the vague notation $\Scr{D}x$ to denote the measure in this
crazy integral.

\begin{figure}
\begin{center}
\begin{tikzpicture}[scale=1.2]
  \pgftext{\includegraphics[width=150pt]{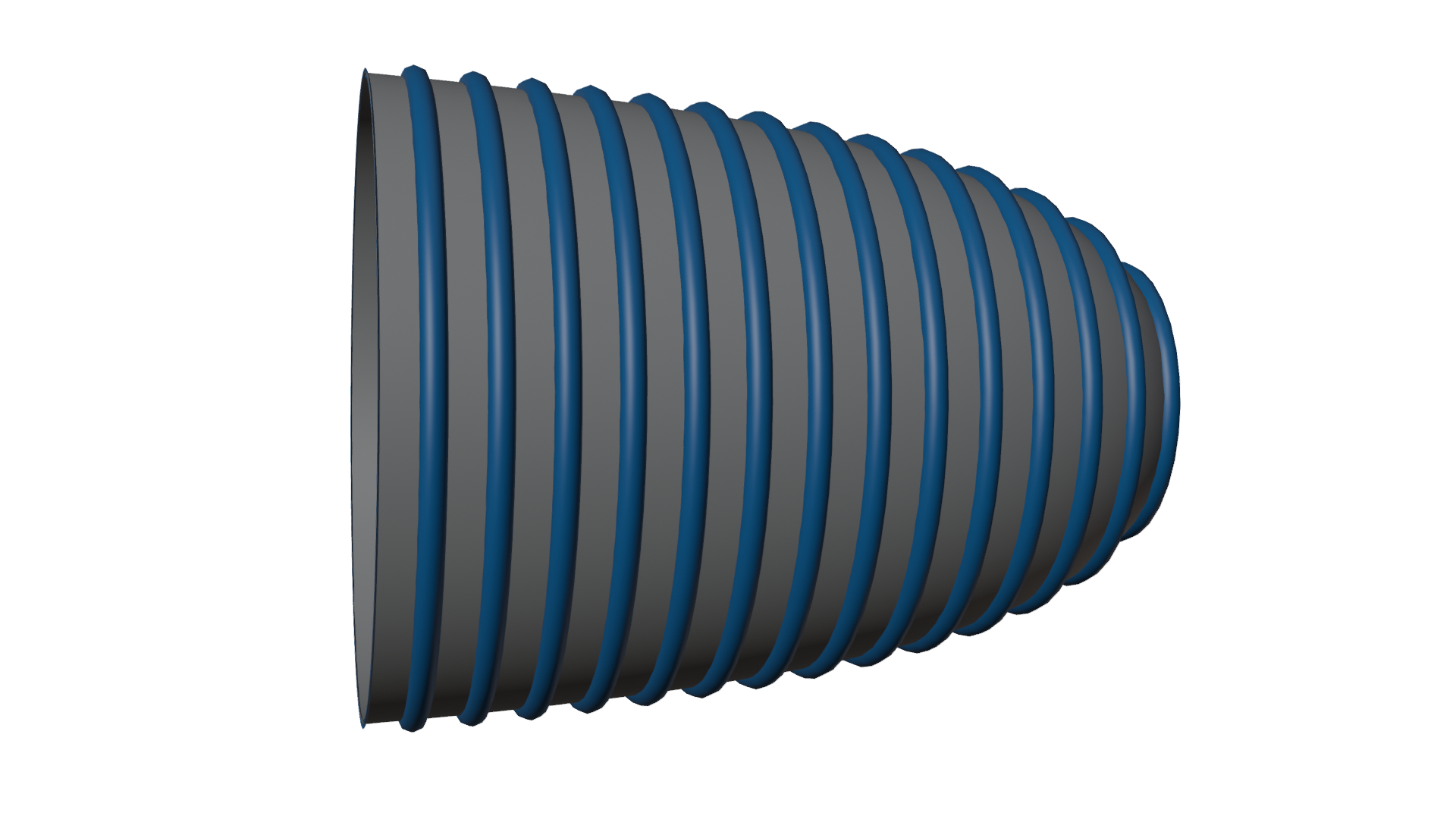}} at (0pt,0pt);
  \draw[->] (3,-2) -- (-2,-2);
  \draw (0,-2) node[anchor=north] {\scriptsize Time};
  \draw (-1.3,-1.45) node {\scriptsize $B$};
  \end{tikzpicture}
\end{center}
\caption{A String Worldsheet.} 
\label{f.worldsheet1}
\end{figure}

We want to replace the notion of a particle with a closed
string. So paths are replaced by cylinders to give something like
figure \ref{f.worldsheet1}. The string wavefunction is
now a map 
\begin{align}
	\psi:L(\mathbb R)\rightarrow \mathbb C,
\end{align}
where $L(\R)$ is the loopspace of $\R$. That is,  a point in $L(\mathbb R)$ is
a map $S^1\rightarrow \mathbb R$. Let's say we're interested in
strings starting out at $t=-\infty$ and then we look at them at time
$t=0$. So we want to impose some boundary condition, $B$, at $t=0$. We decompose this boundary
conditions into Fourier modes. For $t=0$ we put
\begin{align}
	x(e^{i\sigma})=\sum_{n\in \mathbb Z}x_ne^{in\sigma},  \label{eq:xfr}
\end{align}
where $x_{-n}=\bar x_n$ as $x$ is real. So, the path integral gives
\begin{align}
	\psi(x_0,x_1,\dots,x_n,\dots)=\int_B\Scr{D}x\,e^{iS},
\end{align}
where we impose the boundary condition $B$ given by (\ref{eq:xfr}) at
time $t=0$.

This is easier to picture if we use our complex coordinate
$z=e^{i\sigma^+}$. Now $t=-\infty$ is mapped to the origin $z=0$ while
$t=0$ is mapped to the circle $|z|=1$. So our path integral is now
over all {\em disks\/} centered at the origin with some fixed
condition on the boundary of the disk.

The action $S$ is given by the integral of the Lagrangian which is the
kinetic energy {\em minus\/} the potential energy. This turns out to
be given quite nicely as
\begin{align}
	S=\frac 1{4\pi}\int \partial x \bar{\partial}x\,d^2z.
\end{align}
Consider a classical solution $x_{\textrm{cl}}$ which minimizes the
action for a given boundary condition at $t=0$. We can then write
\begin{equation}
  x = x_{\textrm{cl}} + x',
\end{equation}
where $x'$ represents a quantum fluctuation around this classical
value. 

The Euler--Lagrange equations for this action amount to
$\partial\bar\partial x=0$ and so we can write 
\begin{align}
	x_\text{cl}=x_0+\sum_{n>0}(x_nz^n+x_{-n}\bar{z}^n),
\end{align}
where the constants $x_n$ are fixed by the boundary condition. We then
have
\begin{equation}
\begin{split}
S &= \frac1{4\pi}\int \partial x\bar\partial x\,dzd\bar z\\
&= \frac1{4\pi}\int \partial x_{\textrm{cl}}\bar\partial x_{\textrm{cl}}\,dzd\bar z
+ \frac1{4\pi}\int \partial x'\bar\partial x'\,dzd\bar z,
\end{split}
\end{equation}
since minimizing the action removes cross terms.
Integrating over $|z|<1$ we explicitly obtain the classical contribution to
the action
\begin{equation}
\begin{split}
\frac1{4\pi}\int \partial x_{\textrm{cl}}\bar\partial
x_{\textrm{cl}}\,dzd\bar z
&= \frac1{4\pi}\sum_{n>0,m>0} nmx_nx_{-m}\int z^{n-1}\bar z^{m-1}\,dzd\bar z\\
&= \frac1{4\pi}\sum_{n>0,m>0} nmx_nx_{-m}\int_0^{2\pi}\!\!\!\int_0^1
r^{n+m-2}e^{i(n-m)\theta}\,2r\,drd\theta\\
&= \ff12\sum_{n>0} n|x_n|^2,
\end{split}
\end{equation}
where we used $z=re^{i\theta}$.

This implies that
\begin{equation}
  \psi(x_0,x_1,\ldots) = C_0\exp\left(-\ff12\sum_{n>0} n|x_n|^2\right),
  \label{eq:psi1ground}
\end{equation}
where
\begin{equation}
C_0 = \int_{B_0} \mathcal{D}x' \exp\left(-\frac1{4\pi}\int \partial x'\bar\partial x'\,dzd\bar z
\right)
\end{equation}
is the integral over all worldsheets which look like a disk
$|z|\leq 1$ with boundary condition $B_0$ given by $x'=0$ at
$|z|=1$. We have no idea how to compute $C_0$ but it does not depend
on the $x_n$'s so just gives some constant factor in the wave
function.

This is great! We've managed to get the interesting part of the path
integral, that is the part that depends on the boundary condition at
$t=0$, while hiding the nasty part in some normalization we don't
really care about.

So the path integral with boundary modes $x_n$ yields
\begin{align}
	\psi(x_0,x_1,\dots)=C_0 \exp\left(-\ff12
  \sum_{n>0}n|x_n|^2\right). \label{eq:vac1}
\end{align}

But now recall that the wavefunction of a harmonic oscillator is
generated by the basis of Hermite polynomials and is given by
\begin{align}
	\psi(x)=H_n(x)e^{-\omega x^2}.
\end{align}
So we have magically discovered the string wavefunction
(\ref{eq:vac1}) at $t=0$ where every oscillator mode is in its ground
state. This is kind of weird as we never explicitly needed to put any
boundary condition at $t=-\infty$. Somehow the path integral picture
``wants'' to do the right thing and produce the state $\ket0$ at $t=0$
in the absence of any other information.

\begin{figure}
\begin{center}
\begin{tikzpicture}
  \filldraw (0,0) circle (0.05) node[anchor=west] {$\partial x(0)$};
  \draw (0,0) circle (1.5);
  \draw (1.5,1.5) node {$\alpha_{-1}\ket0$};
\end{tikzpicture}
\end{center}
\caption{A vertex operator creates a state at $|z|=1$.} \label{fig:opstate}
\end{figure}
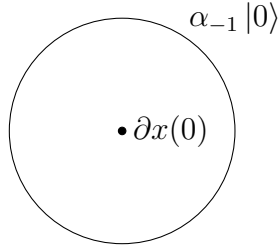

For a bit more magic, let's now stick $\partial x(0)$ into the path
integral:
\begin{equation}
  \int _B \Scr{D}x\,e^{-S}\,\partial x(0).
\end{equation}
But
\begin{equation}
 \partial x_{cl}(0)=x_1,
\end{equation}
and the $x'$ contribution is unaffected. So the path integral
evaluates to
\begin{equation}
  \psi(x_0,x_1,\dots)=C_0 x_1 \exp\left(-\ff12 \sum_n |x_n|^2\right).
\end{equation}
But $x_1$ corresponds to the first Hermite polynomial and so the
lowest mode of the lowest frequency oscillator is excited. This is
$\alpha_{-1}$ excitation. So somehow the function $\partial x(0)$
produced this excitation back at $t=-\infty$. The function $\partial
x(0)$ creates a state as shown in figure \ref{fig:opstate}.

Similarly we may insert
\begin{align}
	\partial^2x(0),
\end{align}
to give the $\alpha_{-2}$ excitation, etc.

\subsection{Vertex Operator Picture}
Notice that we are inserting nothing more than simple {\em
  functions\/} into the path integral to produce new states. We get a
very similar picture by thinking of {\em operators\/} on our Hilbert
space. Let's temporarily use a hat to emphasize operators over functions. So we have
\begin{align}
	i\partial \hat x(z)=\sum_{n\in \mathbb Z}\frac{\alpha_n}{z^{n+1}},
\end{align}
which, when applied to the ground state yields
\begin{align}
	i\partial \hat x(0)\ket 0=\alpha_{-1}\ket{0}.
\end{align}
So the operator $i\partial \hat x(0)$ produced an excitation just like
the function in the path integral did. We will therefore follow the
usual physics convention, confuse these two notions and freely call
functions operators, and, indeed also use the moniker ``fields'' too!

What happens if we instead insert the following operator into the path integral
\begin{align}
	i\partial x(z)?
\end{align}
Since $i\partial x(0)$ produced a state at $z=0$, we will assume the
above ``creates'' this state at some arbitrary point $z$ in the
worldsheet.  We thus produce a kind of two-dimensional quantum field
theory. We can create new things by ``inserting'' functions into the
path integral. Things like $i\partial x(z)$ used this way are often
called ``vertex operators''.

So far we have looked at the past up to some time. To get a complete
story we should continue forward to $t=+\infty$. We thus consider the
future vacuum $\bra0$ to be the adjoint of our past $\ket{0}$.
So we propagate from the past to some boundary condition $B$ at $t=0$
and then forward to the infinite future. This produces an inner product
\begin{align}
  \left\langle 0|0\right\rangle=\int_\text{all possible
  $B$'s}|\psi_1|^2dx_1dx_2\ldots=
  \int_\text{all spheres}\Scr{D}x\,e^{-S}.
\end{align}
We have no idea how to actually compute this but, we can choose to normalize
it by asserting $\left\langle 0|0\right\rangle=1$.

Next we may put some vertex operator into this and compute, for example,
\begin{align}
	\eval{0}{i\partial x(z)}{0}=\eval{0}{\sum_{n\in \mathbb Z}\frac{\alpha_n}{z^{n+1}}}{0}=0,
\end{align}
since every $\alpha_n$ for non-negative $n$ kills $\ket0$ and, for
non-positive $n$, it kills $\bra0$ to the left. Often one takes the
vacuum states as implicit and writes this ``one-point'' function
as
\begin{equation}
  \left\langle\partial x(z)\right\rangle = 0.
\end{equation}

\begin{figure}
\begin{center}
\begin{tikzpicture}
  \draw[dotted] (0,0) circle (0.7);
  \filldraw (30:0.7) circle (0.05) node[anchor=west] {\scriptsize $\partial x(w)$};
  \draw[dotted] (0,0) circle (1.5);
  \filldraw (150:1.5) circle (0.05) node[anchor=west] {\scriptsize $\partial x(z)$};
  \draw (0,0) circle (2);
 \end{tikzpicture}
\end{center}
\caption{A Product of Vertex Operators}  \label{fig:2pt}
\end{figure}
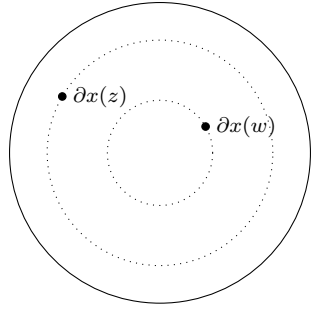

More interestingly, next compute two vertex operators as in figure
\ref{fig:2pt}.
\begin{align}
  \left\langle\partial x(z)\partial x(w)\right\rangle
  &=-\sum_{m>0,n<0}\frac{\eval{0}{\alpha_m\alpha_n}{0}}{z^{m+1}w^{n+1}}\nonumber\\
	&=-\sum_{m>0,n<0}\frac{m\delta _{m+n}}{z^{m+1}w^{n+1}}=-\sum_{m>0}\frac{m}{zw}\left(\frac{w}{z}\right)^m\nonumber\\
	&=-\frac{1}{(z-w)^2},  \label{eq:bbope}
\end{align}
assuming $|w|<|z|$. This assumption is correct as $w$ appears
to the right of $z$ in our path integral and so $w$ happens
``first''. In complex plane language this time ordering is exactly
$|w|<|z|$ as shown in the figure.

\subsection{The Operator Product Expansion}
Note if we write the above product of vertex operators naked outside a path
integral we get
\begin{align}
  \partial \hat x(z)\partial \hat x(w)=-\sum_{m\in \mathbb Z}
  \frac{m}{z w}\left(\frac w z\right)^m,
\end{align}
which {\em nowhere\/} converges. So this product of operators really
does not have literal validity.

So we can either be careful to implicitly ``dress'' our products
carefully inside $\langle\ldots\rangle$ or we can deal with
noncovergent power series by treating them formally. This latter point
of view leads to the mathematical formulation of ``Vertex Operator
Algebras''. We refer to \cite{LL:VOA}, for example, for a full
treatment. We will not pursue this formal language here. Instead we
assume operator products have some kind of intrinsic existence
manifested by what happens when they are put in a path integral.

That said, the above is an example of an ``operator product
expansion'', which might look typically like
\begin{align}
	\mathcal O_i(z)\mathcal O_j(w)=\sum_{n,k}\frac{c_{ij}^{k,n}\mathcal O_k(w)}{(z-w)^k},
\end{align}
for constants $c_{ij}^{k,n}$. Note that in particular cases there
may additionally be terms like $\log(z-w)$ and/or fractional powers.

\begin{figure}
\begin{center}
\begin{tikzpicture}
  \filldraw (0,0) circle (0.05) node[anchor=west] {$w$};
  \begin{scope}[>=stealth,decoration={
    markings,
    mark=at position 0.3 with {\pgftransformscale{2}\arrow{>}}}
    ] 
    \draw[postaction={decorate}] (-0.5,-1) circle (1.6);
  \end{scope}
  \begin{scope}[>=stealth,decoration={
    markings,
    mark=at position 0.3 with {\pgftransformscale{2}\arrow{<}}}
    ] 
    \draw[postaction={decorate}] (-0.7,-1.2) circle (1);
  \end{scope}
  \draw (-0.7,0.32) node {$z$};
  \draw (-0.7,-0.40) node {$z$};
  \filldraw (5,0) circle (0.05) node[anchor=west] {$w$};
  \begin{scope}[>=stealth,decoration={
    markings,
    mark=at position 0.32 with {\pgftransformscale{2}\arrow{>}}}
    ] 
    \draw[postaction={decorate}] (5,0) circle (0.5);
  \end{scope}
  \draw (5.1,0.72) node {$z$};
  \draw[-latex] (2,0) -- +(2,0);
  \end{tikzpicture}
\end{center}
\caption{Change of contours.}
\label{f.contour}
\end{figure}
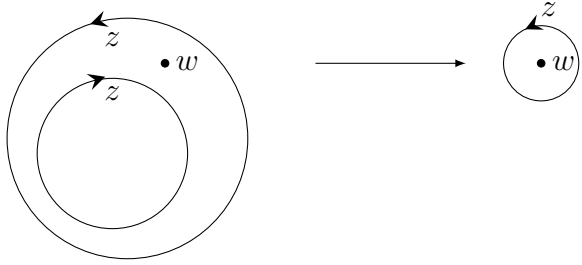

A key fact is that there are relationships between OPE's and
commutation relations. This relates the ``canonical commutator''
picture of quantum mechanics (\ref{eq:xp}) to the path integral.
The order in which we write operators such as $\alpha_n$ becomes time
ordering in the path integral, which, in turn, becomes {\em
  radial\/} ordering in the complex plane. For
example, given the above operator product we can reproduce
$[\alpha_m,\alpha_n]$ using complex analysis as follows. The expansion
(\ref{eq:alphasum}) gives
\begin{equation}
  \alpha_n=\frac 1{2\pi}\aoint _C z^{n}\partial x(z)dz,
\end{equation}
where $C$ encloses the origin once, and thus,
\begin{equation}
  \begin{split}
  [\alpha_m,\alpha_n]&=\frac{1}{4\pi^2}\aoint\aoint_{|w|<|z|}z^mw^n
  \partial x(z)\partial x(w)\,dz\,dw-\,\\ &\hspace{2cm}\frac{1}{4\pi^2}\aoint
                                                \aoint_{|w|>|z|}z^mw^n \partial x(z)\partial x(w)\,dz\,dw,
  \end{split}
\end{equation}
where each integral contour encloses the original once under the
inequalities shown. If we assume the only singularities of the
operator product occur at $z=w$, we may change the $z$-contour for a
fixed value of $w$ as in figure \ref{f.contour}. Using (\ref{eq:bbope}) we recover
\begin{equation}
  \begin{split}
    [\alpha_m,\alpha_n]
    &=-\frac{1}{4\pi^2}\aoint_{w=0}\!\!\!dw \aoint_{z=w}\frac{z^mw^n}{(z-w)^2}\,dz\\
    &=\frac1{2\pi i}\aoint_{w=0}mw^{m+n-1}\,dw\\
    &=m\delta_{m+n}.
\end{split}
\end{equation}

Note that anything nonsingular in the operator product will not
contribute to the contour integrals and so does not affect the
commutator. Therefore, the above calculation only determines the
operator product to leading order:
\begin{align}
	\partial x(z)\partial x(w)=-\frac{1}{(z-w)^2}+\dots,
\end{align}
where the ellipses denote something finite as $z\rightarrow w$. Such
use of ellipses is standard in operator product notation. Note,
however, that these finite terms can still be important. If we compute a
correlation function with more than two vertex operators, finite
terms can lead to singular ones by the further operator products. This
all adds to the complications of the vertex operator picture.

We can do the analogue of normal ordering for operator products:
\begin{align}
	\normord{\mathcal O(z)\mathcal O(w)}\,=\mathcal O(z)\mathcal O(w)-(\text{singular poles as $z\rightarrow w$}),
\end{align}
and define
\begin{equation}
  \normord{\mathcal O(z)\mathcal O(z)}\,=
  \lim_{z\to w} \normord{\mathcal O(z)\mathcal O(w)}.
\end{equation}
Notice we haven't used the unadorned field $x(z)$ itself as a vertex
operator. While it can be, the logarithm in (\ref{eq:x}) produces
cuts that make it awkward. Differentiating rids us of this, but
we can also exponentiate. This implies it is interesting to consider
\begin{align}
	V(z) = \normord{e^{ikx(z)}}.
\end{align}
Indeed, this is the vertex operator that creates a string state with center of
mass momentum $p=k$. We'll implicitly assume normal ordering in this
operator from now on. One can derive the following products:
\begin{equation}
  \begin{split}
  e^{ikx(z)}\,e^{ilx(z)} &=
          (z-w)^{kl}e^{ikx(z)+ilx(w)}\\
  i\partial x(z)e^{ikx(w)} &= \frac{k}{z-w}e^{ikx(z)}+\ldots,
  \end{split}  \label{eq:pOPE}
\end{equation}
which we will need later.

Another operator of note is $T(z)$ whose Fourier modes are built out
of the $L_m$'s.
\begin{align}
	T(z)=\sum_{m\in \mathbb Z}\frac{L_m}{z^{m+2}}.
\end{align}
This is known as the ``stress (energy) tensor''.

The Virasoro algebra then becomes
\begin{align}
	T(z)T(w)=\frac{c/2}{(z-w)^4}+\frac{2T(w)}{(z-w)^2}+\frac{\partial
  T(w)}{z-w}+\dots  \label{eq:TODE}
\end{align}

\subsection{Conformal Field Theory}

We can now sort of give a very rough outline of a ``definition'' of conformal field theory.

The data is a collection of vertex operators $\mathcal O_1$,
$\mathcal O_2$,\ldots, together with {\em correlation functions\/}
computed as follows. Given a fixed Riemann Surface $\Sigma$ we
can insert any vertex operators at points on this surface and compute
the path integral
\begin{align}
  \langle\mathcal O_1(z_1)\mathcal O_2(z_2)\dots
  \mathcal O_j(z_j)\dots)\rangle_\Sigma\in \mathbb C.
\end{align}

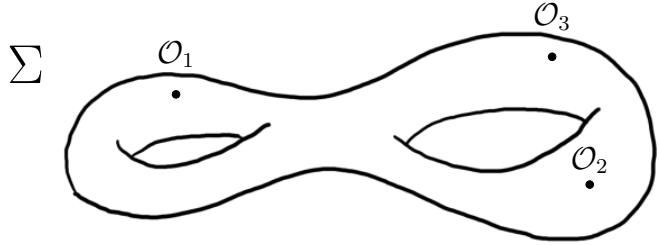
\begin{figure}
\begin{center}
  \begin{tikzpicture}
    \pgftext{\includegraphics[width=250pt]{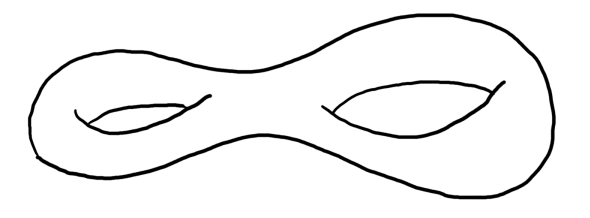}} at (0pt,0pt);
    \filldraw (-2.5,0.6) circle (0.05) node[anchor=south,inner
    sep=10pt]
    {$\mathcal{O}_1$};
    \filldraw (3,-0.6) circle (0.05) node[anchor=south,inner
    sep=4pt] {$\mathcal{O}_2$};
    \filldraw (2.5,1.1) circle (0.05) node[anchor=south,inner
    sep=10pt] {$\mathcal{O}_3$};
    \draw (-4.5,1) node {\LARGE $\Sigma$};
\end{tikzpicture}
\end{center}
\caption{A Riemann surface with vertex operators at marked points.}
\label{f.riemann}
\end{figure}

One of the vertex operators in our collection is $T(z)$. Everything
must then be invariant under the action of the resulting conformal
symmetry. For example, $L_{-1}$ must correctly generate translations
leading to ``Ward'' identities. Also, if $\Sigma$ is a sphere then all
correlation functions are invariant under the M\"obius group. An
immediate consequence, for example, is that a suitably normalized self-adjoint
vertex operator will always have a product
\begin{equation}
  \mathcal O(z)\mathcal O(w) = \frac1{(z-w)^{2h}}+\ldots,  \label{eq:spinstat}
  \end{equation}
where $h$ is the weight of $\mathcal O$. All of
this is carefully explained in \cite{Gins:lect} and we need not
reproduce it here. 

Another consideration is that Riemann surfaces can degenerate into
connected sums of other Riemann surfaces. Thus one can make statements
about how the correlation functions vary as we move in the moduli
space of (marked) Riemann surface. Again, we need not pursue this
here.

Finally, of course, we have only talked about the {\em holomorphic\/}
dependence of $z$ and have thus completely ignored the right-moving
string modes. To get to final answers one needs to assemble the
holomorphic and antiholomorphic parts together.

A correlation function of particular interest is that of a genus one Riemann
surface with no operators inserted. This is a function of the complex
structure of the Riemann surface alone and is thus a relatively simple
object. It is called the ``partition function''. Again, a sensible
course on conformal field theory would have much to say about this but
we need to move on.

\section{Day 3}

\subsection{Fermions}
Everything we need to know about the ``bosonic'' oscillator modes of
our string we have considered so far can be derived from the commutator
\begin{align}
	[\alpha_m,\alpha_n]=m\delta_{m+n}.
\end{align}

We'd now like to consider some new kind of degree of freedom our
string may exhibit. We'll denote its modes by $a_r$.  So let's try the
simplest thing one might think of:
\begin{align}
 [a_r,a_s]=\delta_{r+s}.  \label{eq:acom}
\end{align}
Now, if we exchange $r$ and $s$ the right-hand-side of this equation
is invariant. So this cannot be a commutator. Instead, the obvious
thing to do is to declare it to be an {\em anticommutator}:
\begin{equation}
  [a_r,a_s] = a_ra_s+a_sa_r.
\end{equation}
We thus say that these $a_r$ excitations correspond to a {\em
  fermion}. Some people like using braces for anticommutators but one
can write general equations more easily if you use the convention that
$[A,B]$ is a commutator unless both $A$ and $B$ are fermions in which
case it is an anticommutator. 

If we assume $a_r$ comes from a primary field we can find its weight by
computing the Jacobi identity (with the correct signs allowing for fermions):
\begin{align}
	[L_m,[a_r,a_s]]-[a_s,[L_m,a_r]]+[a_r,[a_s,L_m]]=0.
\end{align}
From (\ref{eq:acom}) and (\ref{eq:prim-com}) this yields
\begin{equation}
  m(2h-1)\delta_{m+r+s}=0,
\end{equation}
which fixes $h=\ff12$. So we define our new fermion vertex operator
as
\begin{equation}
  \psi(z)=\sum_r \frac{a_r}{z^{r+1/2}},  \label{eq:psidef}
\end{equation}
and we get an operator product
\begin{equation}
	\psi(z)\psi(w)=\frac{1}{z-w}+\text{finite as $z\rightarrow w$}.
\end{equation}
  
It's fun to note we have a very simple picture of a two-dimensional
spin-statistics theorem at work here. We see from (\ref{eq:spinstat})
that exchanging vertex operators changes sign or not whether $h$ is in
$\Z$ or $\Z+\ff12$. Consistent with this, the bosons $i\partial x(z)$
have $h=1$ and these fermions have $h=\ff12$.

Now, we see from (\ref{eq:psidef}) that either $r\in\Z$ and $\psi(z)$
has a branch cut at the origin, or $\psi(z)$ is single-valued and
$r\in\Z+\ff12$. So, which is more natural? It turns out that we need
to consider both possibilities, leading to two {\em sectors\/} of
fermions. They are named as
\begin{align}
	r\in\begin{cases}
		\mathbb Z+\ff12&\text{ Neveu Schwarz (NS)}\\
		\mathbb Z&\text{ Ramond (R)}.
	\end{cases}
\end{align}

Following the process we did earlier for the boson, we find Virasoro
generators giving the correct commutation relations with $a_r$ from
\begin{equation}
  L_n = \ff12\sum_rr\normord{a_{-r}a_{n+r}},  \label{eq:Lfermion}
\end{equation}
which is equivalent to
\begin{equation}
  T(z) = -\ff12\normord{\psi(z)\partial\psi(z)}.
\end{equation}
Again, a careful computation shows that this then obeys the Virasoro
algebra, but this time the central charge is $c=\ff12$, in either
sector. (This works perfectly for the NS sector but for the Ramond
sector we also see a shift in $L_0$ by $c/24$.) So these fermions are
somehow ``worth'' one half of a boson!

To construct the spectrum, we first need a vacuum. There is a stark
difference between the NS vacuum and the R vacuum which will be
absolutely key for us.

The NS vacuum is easy enough to define since $r$ is not integral:
\begin{align}
	a_r\ket 0_\text{NS}=0\text{ for }r>0.
\end{align}
We can declare this to be the unique NS vacuum and the lowest excited state
is $a_{-\frac12}\ket0_\text{NS}$, etc.

For the Ramond sector we need to work harder because $a_0$ does not
change the energy. Consider $D$ dimensions by using operators
$a_r^\mu$, for $\mu=1,\ldots,D$. We have
\begin{align}
	[a_0^\mu,a_0^\nu]=\delta^{\mu,\nu}.
\end{align}
This is well-known in mathematics as a {\em Clifford Algebra\/} on
$\mathbb R^D$ (with standard inner product).

We cannot say $a^\mu_0$ kills the vacuum as this is inconsistent with
$(a_0^\mu)^2=\ff12$, and we cannot say $a^\mu_0$ leaves the vacuum
invariant (or rescales it) as that is inconsistent with the Clifford
algebra for $\mu\neq\nu$. So the Ramond vacuum must be
multidimensional (assuming $D>1$). 

We can build things like
\begin{align}
	\dots a_0^{\mu_3}a_0^{\mu_2}a_0^{\mu_1}\ket 0_R,
\end{align}
choosing to include any $a^\mu_0$ or not and these vectors in the
Hilbert space all have the same energy. If all such possibilities are
considered linearly independent we would construct a $2^D$-dimensional
vacuum, but this turns out to be rather redundant. To be precise, it
is {\em reducible}. There is only one irreducible representation of
the Clifford algebra (assuming $D$ is even) and that corresponds to a
Dirac spinor of dimension $2^{\frac D2}$ (see, for example,
\cite{FulHar:rep}). As such, the Ramond vacuum corresponds to a
spacetime {\em spinor}. In contrast the NS vacuum is a spacetime {\em
  scalar\/} whilst the NS states
\begin{equation}
  a_{-\frac12}^\mu\ket0_{\textrm{NS}},
\end{equation}
form a spacetime {\em vector}.

It is essential here to keep track of whether we are talking about the
worldsheet or spacetime when it comes to bosons and fermions. We have
bosons $\partial x(z)$ and fermions $\psi(z)$ on the worldsheet,
whereas spacetime bosons are in the NS sector and spacetime fermions
(i.e., spinors) are in the R sector.

\subsection{Supersymmetry}
Let's have both $\alpha_n^\mu$ and $a_r^\mu$ degrees of freedom for
our $D$-dimensional string. So the Virasoro generators are the sum of
(\ref{eq:Lboson}) and (\ref{eq:Lfermion}). Define the fermionic object
\begin{align}
	G_r=\sum_{n,\mu} \alpha^\mu_{-n}a^\mu_{n+r},
\end{align}
(which may be in the NS or R sector). Then (grading the brackets
correctly)
\begin{align}
	[G_r,\alpha_m^\mu]=-ma_{r+m}^\mu,\qquad
	[G_r,a_s^\mu]=\alpha_{r+s}^\mu.
\end{align}
This is therefore a ``worldsheet supersymmetry''. It exchanges
worldsheet fermions with worldsheet bosons.

One can then show (in the NS sector) that this obeys the ``$N=1$
superconformal algebra" defined by
\begin{equation}
  \begin{split}
	[L_m,L_n]&=(m-n)L_{m+n}+\frac{c}{12}m(m^2-1)\delta_{m+n}\\
	[L_m,G_r]&=\left(\ff12m - r\right) G_{m+r}\\
	[G_r,G_s]&=2L_{r+s}+\frac{c}{3}\left(r^2-\ff14\right)\delta_{r+s},
  \end{split}
\end{equation}
with $c=\ff32 D$. (Again, that last term in the third equation is a
good exercise in treating normal ordering correctly.) The $N=1$ refers
to the fact we only have one $G$. We'll get a second one shortly.
Alternatively, introducing the ``supercurrent''
\begin{equation}
  G(z)=\sum_r\frac{G_r}{z^{r+3/2}},
\end{equation}
this algebra can be written as an operator product statement as
(\ref{eq:TODE}) together with
\begin{equation}
  \begin{split}
    T(z)G(w)&=\frac{\frac{1}{2}G(w)}{(z-w)^2}+\frac{\partial G(w)}{z-w}+\dots\\
    G(z)G(w)&=\frac{\ff23 c}{(z-w)^3}+\frac{2T(w)}{z-w}+\dots.
\end{split}
\end{equation}

If we merely added these fermionic degrees of freedom to our
reparametrization-invariant string theory we'd need $\ff32D=26$, which
we can't solve for integral $D$. Instead we define the {\em
  superstring\/} as the model having the $N=1$ superconformal
algebra as a {\em gauge symmetry}. So, we have some kind of
super-reparametrization invariance. This requires adding in
``superghosts'' to the BRST process which raises a whole new
collection of subtleties.

The superghosts themselves have a central charge of 11 so now we
require
\begin{equation}
  \ff32D - 26 + 11=0,
\end{equation}
and we obtain the well-known result that the superstring likes to
live in 10 dimensions. Modular invariance and a peculiar fact about
the ground state of the superghosts force a thing called the GSO
projection to be taken onto odd fermion numbers and that very nicely
removes the offending tachyon we noted earlier. The full story for
this doesn't appear until volume 2 of Polchinski
\cite{Polchinski:1998rr} and so we certainly do not have enough time
to dig deeper into this here.

\subsection{Compactification}

If we wanted to claim that the universe is described by superstring
theory we need to explain away this issue of 10 dimensions. If $D$
were equal to 4 our net central charge would be $-9$. So what we need
to do is to introduce more ``stuff'' into our conformal field theory
to add 9 to $c$. Assuming everything factors nicely, we can treat this
stuff as its own $c=9$ conformal field theory.

So what we need to describe a ``realistic'' superstring theory is some
$c=9$ conformal field theory. One way to interpret this theory is as
providing the 6 ``missing dimensions''. In other words, ten-dimensional
spacetime is viewed as four-dimensional spacetime times some
six-dimensional space $X$ that is ``compact'' enough that we wouldn't
notice it. As such this process is called ``compactification''.

It should be emphasized that at no time did string theory demand that
our $c=9$ conformal field theory have any kind of geometric
interpretation. That said, all the fruitful applications of string
theory to geometry are based on this assumption.

From now on we define $d$ as follows. Let us say that the conformal
field theory we are using to compactify has central charge
$c=3d$. This means we are compactifying down to $10-2d$ observable
non-compactified spacetime and any putative compactification space $X$
is of (real) dimensional $2d$.

\subsection{Spacetime Supersymmetry}

As we've described things so far, we have one vacuum $\ket
0_\text{NS}$ for the NS sector and and a whole subspace of vacua $\ket
0_\text{R}$ for the R sector. We can describe this more succinctly
using the vertex operator language. Let {\em the\/} vacuum $\ket0$ be
the NS vacuum. This is, after all, unique. Then suppose we have a
vertex operator $\Sigma^+(z)$ that creates a direction in the R-vacuum
from this.

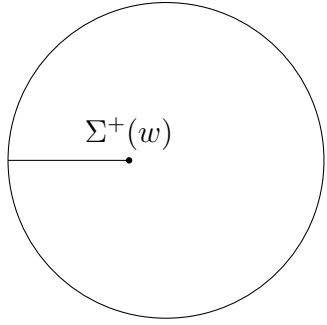
\begin{figure}
\begin{center}
\begin{tikzpicture}[scale=0.7]
  \draw (0,0) circle (3);
  \draw (-3,0) -- (-0.7,0);
  \filldraw (-0.7,0) circle (0.05) node[anchor=south] {$\Sigma^+(w)$};
  \end{tikzpicture}
\end{center}
\caption{A vertex operator creates a branch cut.}
\label{f.branchcut}
\end{figure}

This means that $\Sigma^+(z)$ must create a branch cut in the complex
plane, as in figure \ref{f.branchcut}, so that fermions pick up a sign
as we cross the cut and bosons do not. Our operator product with
fermions must therefore contain square roots, something like
\begin{align}
	\psi^\mu(z)\Sigma^+(w)=\frac{\Xi(w)}{\sqrt{z-w}}+\dots,
\end{align}
whereas the operator product between $\Sigma^+$ and bosons should be
single-valued. 

This operator $\Sigma^+$ transforms the NS sector to the R sector. As
such, it takes spacetime bosons to fermions and so is a {\em spacetime
supersymmetry}.

Let $\Sigma^-$ be the adjoint of $\Sigma^+$. We then claim
\begin{align}
	\Sigma^+(z)\Sigma^{-}(w)=\frac{1}{(z-w)^{d/4}}+\frac{J(w)}{(z-w)^{d/4-1}}+\hbox{less singular},
\end{align}
where $J(w)$ is an $h=1$ vertex operator. This is proven as
follows. That shift in the $L_n$ commutation relations we noted
earlier for the R vacuum means $\Sigma^\pm$ has conformal weight
$d/8$.  That accounts for the first term above. We then argue that the
branch cuts in the product must be consistent so that we get
single-valued correlation functions at the end of the day. This
implies all powers of $z-w$ must differ by an integer. This yields the
second term so long as we can show that term in nonzero. That can be
shown by looking at 4-point functions of $\Sigma^\pm$
\cite{Banks:1987cy}. Furthermore, this 4-point function yields
\begin{equation}
  J(z)J(w) = \frac{d}{(z-w)^2} +\ldots.
\end{equation}

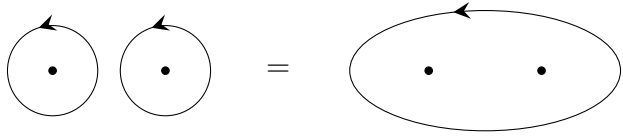
\begin{figure}
\begin{center}
  \begin{tikzpicture}
    \begin{scope}[>=stealth,decoration={
        markings,
        mark=at position 0.3 with {\pgftransformscale{2}\arrow{>}}}
      ] 
      \filldraw (0,0) circle (0.05);
      \draw[postaction={decorate}]  (0,0) circle (0.6);
      \filldraw (1.5,0) circle (0.05);
      \draw[postaction={decorate}]  (1.5,0) circle (0.6);
      \filldraw (5,0) circle (0.05);
      \filldraw (6.5,0) circle (0.05);
      \draw[postaction={decorate}]  (5.75,0) ellipse (1.8 and 0.8);
      \draw (3,0) node {$=$}; 
    \end{scope}
  \end{tikzpicture}
\end{center}
\caption{Conservation of Charge.}
\label{fig:conserve}
\end{figure}

So, to recap here, the existence of spacetime supersymmetry has
required us to have a new field with $h=1$: 
\begin{align}
	J(z)=\sum_n\frac{J_n}{z^{n+1}}.
\end{align}
Any $h=1$ field is a ``current'' in the sense that it has a conserved change
\begin{align}
	J_0=\frac{1}{2\pi i}\aoint_C J(z)dz,
\end{align}
so that if a field $A(z)$ obeys
\begin{equation}
  J(z)A(w) = \frac{q(A)A(w)}{z-w} +\ldots,
\end{equation}
then $A(z)$ is said to have charge $q(A)$, and interactions will
conserve charges by a simple contour argument as in figure
\ref{fig:conserve}. (Our weight one field $i\partial x(z)$ has a
conserved charge given by center of mass momentum.)

This new charge, measured by $J_0$, is called ``$R$-charge''.

Now we have a new field $J(z)$ we should add it to our super-Virasoro
algebra. The only ingredient we are missing is the charge of $G(z)$
with respect to our new current. One can argue
\begin{align}
	G(z)=\ff{1}{\sqrt 2}\left(G^+(z)+G^-(z)\right),
\end{align}
where $G^\pm$ has $R$-charge $\pm 1$.

All in all, this gives a new algebra with generators $T(z)$, $G^\pm(z)$
and $J(z)$ and operator products
\begin{equation}
  \begin{split}
    T(z)T(w) &=\frac{\ff32d}{(z-w)^4}+\frac{2T(w)}{(z-w)^2}+\frac{\partial
               T(w)}{z-w}+\dots\\
    T(z)G^\pm(w)&=\frac{\ff32G^\pm(w)}{(z-w)^2}+\frac{\partial
              G^\pm(w)}{z-w}+\dots\\
	T(z)J(w)&=\frac{J(w)}{(z-w)^2}+\frac{\partial J(w)}{z-w}+\dots\\
	J(z)G^\pm(w)&=\pm\frac{G^\pm(w)}{z-w}+\dots \\
	J(z)J(w)&=\frac{d}{(z-w)^2}+\dots \\
    G^+(z)G^+(w) &= G^-(z)G^-(w) = 0\\
    G^+(z)G^-(w)&=\frac {2d}{(z-w)^3}+
                  \frac{2J(w)}{(z-w)^2}+\frac{2T(w)+\partial J(w)}{z-w}+\dots.
\end{split}  \label{eq:N=2OPE}
\end{equation}
This is called the ``$N=2$ superconformal algebra''. So spacetime
supersymmetry implies this $N=2$ algebra. If we demand a similar
condition in the right-moving sector we get an antiholomorphic copy of
this generated by $\tilde T(z)$, $\tilde G^\pm(z)$
and $\tilde J(z)$ to give an $N=(2,2)$ superconformal field theory.

It is important to note that we do {\em not\/} gauge this $N=2$
superconformal algebra. No further ghosts or superghosts are added to
the $N=1$ system and so no central charge conditions are modified.

We can play a useful trick with vertex operator algebras to understand
the $R$ charges more. Let's assume we can integrate $J(z)$ to form a
new boson $\phi(z)$. If this boson behaves just like $x$ we need to
set
\begin{equation}
  J(z) = i\sqrt{d}\partial\phi(z),
\end{equation}
to get normalizations right. It follows from (\ref{eq:pOPE}) that any
field of the form
\begin{equation}
  A(z)=e^{\frac{ij}{\sqrt d}\phi(z)}A'(z),
\end{equation}
where $A'(z)$ has no $\phi$-dependence, has $R$-charge equal to
$j$. One may choose
\begin{equation}
  \Sigma^\pm(z) = e^{\pm\frac i2\sqrt d\phi(z)},
\end{equation}
and so, again from (\ref{eq:pOPE}), it follows that
\begin{equation}
  \Sigma^+(z)A(w) = (z-w)^{\frac j2}\ldots.
\end{equation}
So, the rule for how the operator $\Sigma^+$ makes branch cuts with
respect to other operators becomes
\begin{equation}
  j = \begin{cases} \textrm{even}&\textrm{for bosons}\\
    \textrm{odd}&\textrm{for fermions}\\
  \end{cases}
\end{equation}

In other words, the $\Z_2$-grading of fermion number is extended to a
$\Z$-grading for the $R$-charge. So spacetime supersymmetry requires
both an $N=2$ superconformal algebra, together with the above
$R$-charge quantization condition.

The fact that $R$-charge gives this $\Z$-grading is a profound statement in many
modern applications of conformal field theory and string theory to
geometry. If there is some kind of $\Z$-grading (coming from, say, a
differential graded algebra) it ultimately comes from something like
the spacetime supersymmetry condition. Without it, you probably only
have a $\Z_2$-grading.

\section{Day 4}

\subsection{Chiral Primary Fields}

Consider an $N=(2,2)$ superconformal field theory producing spacetime
supersymmetry as we discussed in the previous lecture.  The last of
the products in (\ref{eq:pOPE}) can be written
\begin{equation}
  [G_r^+,G_s^-] = 2L_{r+s} + (r-s)J_{r+s} + d(r^2-\ff14)\delta_{r+s}.
\end{equation}
Suppose $\ket\psi$ has eigenvalues $h$ and $j$ for $L_0$ and $J_0$
respectively. Then
\begin{equation}
  \begin{split}
    \eval{\psi}{[G_{1/2}^-,G_{-1/2}^+]}{\psi }&=\left\|G_{-1/2}^+\ket
                     \psi\right\|^2+\left\|G_{1/2}^-\ket \psi\right\|^2\\
	&=(2h-j)\left\|\ket \psi\right\|^2.
\end{split}
\end{equation}
So $j\leq 2h$ with equality if and only if
\begin{align}
	G_{1/2}^-\ket \psi = G_{-1/2}^+\ket \psi=0.
\end{align}
This is called a ``chiral primary field'' \cite{LVW:}. We may
similarly prove $-j\leq h$ with equality when
\begin{align}
	G_{1/2}^+\ket \psi = G_{-1/2}^-\ket \psi=0.
\end{align}
defining an ``antichiral primary field''. Furthermore, by considering
the case $r=\ff32$ it is easy to show
\begin{equation}
  -d\leq j \leq d.
\end{equation}
The same conditions hold in the right-moving sector with the same
statements for $\bar h$ and $\bar\jmath$. Fields which are chiral
primary in both sectors are called (c,c)-fields, etc.

Let $k^{j,\bar\jmath}$ be the dimension of the space of (c,c) states with
eigenvalues of $J_0$, $\tilde J_0$ equal to
$j, \bar\jmath$ respectively. Note we have already shown that $j$ and $\bar\jmath$
are integers, so we have a finite set of numbers. It is traditional to
arrange them in a diamond:
\begin{equation}
  \setlength\arraycolsep{6pt}
  \begin{eqmatrix}[10pt]
    &&&&k^{0,0}&&&&\\
    &&&k^{1,0}&&k^{0,1}&&&\\
    &&k^{2,0}&&k^{1,1}&&k^{0,2}&&\\
    &&&&\vdots&&&&\\
    k^{d,0}&&k^{d-1,1}&&\cdots&&&&k^{0,d}\\
    &&&&\vdots&&&&\\
    &&&~k^{d,d-1}&&k^{d-1,d}&&&\\
    &&&&k^{d,d}&&&&\\
\end{eqmatrix}  \label{eq:cc-diamond}
\end{equation}

The product of $\Sigma^+(z)$ with itself produces a useful field
\begin{equation}
  \Omega(z) = e^{i\sqrt{d}\phi(z)},
\end{equation}
which itself is chiral primary of charge $d$. Furthermore it is the
only such chiral primary field. It follows that $d\in\Z$, and we
determine the corners of our diamond:
\begin{equation}
  k^{0,0}=k^{d,0}=k^{0,d}=k^{d,d} = 1.
\end{equation}

The product of $\Omega(z)$ with an antichiral primary field of charge
$-j$ yields a chiral primary field of charge $d-j$. The adjoint of a
(c,c)-field of charge $(j,\bar j)$ is an (a,a)-field of charge
$(-j,-\bar\jmath)$ and we can then use $\Omega(z)$ and
$\bar\Omega(\bar z)$ to map this to a (c,c)-field of charge
$(d-j,d-\bar\jmath)$. This is an invertible process. It follows that
\begin{equation}
  k^{j,\bar\jmath} = k^{d-j,d-\bar\jmath}.
\end{equation}

If we have some kind of holomorphic-antiholomorphic symmetry on the
worldsheet, then it might follow that
$k^{j,\bar\jmath}=k^{\bar\jmath,j}$. Then our diamond has
left-right and up-down symmetry. This is almost (but not quite) always
the case in any example in the literature.

\subsection{The Torus}
The simplest case is $d=1$ and then our diamond is completely
determined:
\begin{equation}
 \setlength\arraycolsep{2pt}
  \begin{eqmatrix}[10pt]
    &1&\\
    1&&1\\
    &1&\\
\end{eqmatrix}.  \label{eq:1torusHD}
\end{equation}

We can construct this conformal field theory as follows.  Take two
bosons $x^{1,2}$ and two fermions $\psi^{1,2}$ (plus
$\tilde \psi^{1,2}$ for right movers).

Define
\begin{align}
	x^\pm=\ff1{\sqrt{2}}(x^1\pm i x^2),\qquad
	\psi^\pm=\ff{1}{\sqrt 2}(\psi^1\pm i \psi^2).
\end{align}
That is, we use complex coordinates on the {\em target space\/} now as
well as the worldsheet. If we impose periodic conditions, we can thus
view $x^+$ as the coordinate on a complex torus of complex dimension
1.

Now we obtain our desired $N=(2,2)$ superconformal field theory with
\begin{equation}
  \begin{split}
    T(z)&=-\partial x^+(z)\partial x^-(z)-
          \ff12\psi^+(z)\partial \psi^-(z)-\ff12 \psi^-(z)\partial \psi^+(z)\\
	G^\pm(z)&=i\sqrt 2 \partial x^\mp(z) \psi^\pm(z)\\
	J(z)&=\psi^+(z)\psi^-(z),
\end{split}
\end{equation}
and similarly for the antiholomorphic sector.

Note that $\psi^+(z)$ gives $k^{1,0}=1$ and $\tilde\psi^+(z)$ gives
$k^{0,1}=1$. In this case $\Omega(z)=\psi^+(z)$.

To get a higher value of $d$, we can simply take $d$ copies of
this. The cleanest notation for this is to use $x^i$, $i=1,\ldots,d$
for the holomorphic coordinates and $x^{\bar\imath}$,
$\bar\imath=1,\ldots,d$ for the antiholomorphic coordinates. Again, we
need to emphasize the distinction between worldsheet and target
space. Complex conjugation on the worldsheet would take $\partial x^i$
to $\bar\partial x^i$, whereas complex conjugation in the target space
takes $\partial x^i$ to $\partial x^{\bar\imath}$!

We can now build (c,c) fields by taking combinations of $\psi^i$'s and
$\tilde\psi^i$'s (which anticommute but otherwise do not interact with
each other). For example, taking 3 copies of our complex torus we get
$d=3$ and a (c,c)-diamond
 \begin{equation}
 \setlength\arraycolsep{2pt}
  \begin{eqmatrix}[10pt]
    &&&1&&&\\
    &&3&&3&&\\
    &3&&9&&3&\\
    1&&9&&9&&1\\
    &3&&9&&3&\\
    &&3&&3&&\\
    &&&1&&&\\
\end{eqmatrix}  \label{eq:3torusHD}
\end{equation}

\subsection{The Hodge Diamond}

Anyone familiar with complex geometry will instantly recognize
(\ref{eq:3torusHD}) as the ``Hodge Diamond'' of the complex
3-torus. Let us quickly review what this is.  A complex manifold $X$ has
local homolorphic coordinates $x^i$ and antiholomorphic coordinates
$x^{\bar\imath}$.  A differential form of type $(p,q)$ is of the form
\begin{equation}
  \omega = f(x^i,x^{\bar\imath}) dx^{i_1}dx^{i_2}\ldots
  dx^{i_p}dx^{\bar\imath_1}dx^{\bar\imath_1}\ldots dx^{\bar\imath_q},
\end{equation}
where the differentials $dx$ anticommute. Let $\Omega^{p,q}$ be the
space of all such forms. We can then form a double complex
\begin{equation}
  \begin{xy}
  \xymatrix{
    \vdots&\vdots&\vdots&\\
    \Omega^{0,2}\ar[r]^{\partiald}\ar[u]^{\bar\partiald}&\Omega^{1,2}\ar[r]^{\partiald}\ar[u]^{\bar\partiald}&\Omega^{2,2}\ar[r]^{\partiald}\ar[u]^{\bar\partiald}
    &\cdots\\
    \Omega^{0,1}\ar[r]^{\partiald}\ar[u]^{\bar\partiald}&\Omega^{1,1}\ar[r]^{\partiald}\ar[u]^{\bar\partiald}&\Omega^{2,1}\ar[r]^{\partiald}\ar[u]^{\bar\partiald}
    &\cdots\\
    \Omega^{0,0}\ar[r]^{\partiald}\ar[u]^{\bar\partiald}&\Omega^{1,0}\ar[r]^{\partiald}\ar[u]^{\bar\partiald}&\Omega^{2,0}\ar[r]^{\partiald}\ar[u]^{\bar\partiald}
    &\cdots\\
  }
\save="x"!LD+<-3mm,0pt>;"x"!RD+<10pt,0pt>**\dir{-}?>*\dir{>}\restore
\save="x"!LD+<0mm,-3mm>;"x"!LU+<0mm,2mm>**\dir{-}?>*\dir{>}\restore
\save!RD+<6mm,0mm>*{p}\restore
\save!LU+<-2mm,-4mm>*{q}\restore
\end{xy}  \label{eq:FroHss}
\end{equation}
with differentials
\begin{equation}
  \begin{split}
    \partiald\omega &= \sum_j\frac{\partial
    f(x^i,x^{\bar\imath})}{\partial x^j} dx^jdx^{i_1}dx^{i_2}\ldots dx^{i_p}dx^{\bar\imath_1}dx^{\bar\imath_1}\ldots dx^{\bar\imath_q}\\
    \bar\partiald\omega &= \sum_{\bar\jmath}\frac{\partial
    f(x^i,x^{\bar\imath})}{\partial x^{\bar\jmath}} dx^{\bar\jmath}dx^{i_1}dx^{i_2}\ldots dx^{i_p}dx^{\bar\imath_1}dx^{\bar\imath_1}\ldots dx^{\bar\imath_q}.
  \end{split} \label{eq:ds}
\end{equation}
By a construction called the Fr\"olicher spectral sequence, this
computes the de Rham cohomology of $X$ with respect to a total
differential
\begin{equation}
  d = \partiald + \bar\partiald,
\end{equation}
with a filtration that determines the {\em Hodge Numbers\/} $h^{p,q}$
as the dimension of the spaces in this grid after applying cohomology
with respect to $\partiald$ and $\bar\partiald$ (i.e., the $E_2$ stage
of the spectral sequence, where it degenerates). We refer to
\cite{Voisin:H1} for a nice account of this.

So the question is, can we relate our (c,c)-dimensions
$k^{j,\bar\jmath}$ to the Hodge numbers $h^{p,q}$? The first thing to
note is that the two superscripts in $k^{j,\bar\jmath}$ are
swapped by worldsheet complex conjugation whilst the two superscripts
in $h^{p,q}$ are swapped by target space complex conjugation, and we
have emphasized these are not the same thing! So we need to combine
them somehow. The trick is to consider (c,a) fields of the form
\begin{equation}
  A(z,\bar z) = f(x^i,x^{\bar\imath}) \psi^{i_1}\psi^{i_2}\ldots
  \psi^{i_p}\tilde\psi^{\bar\imath_1}\tilde\psi^{\bar\imath_1}\ldots \tilde\psi^{\bar\imath_q},
\end{equation}
Now, if we define the worldsheet supersymmetry action
\begin{equation}
  Q^+\mathcal O(w)=\frac1{2\sqrt2\pi}\aoint_{z=w} G^+(z)\mathcal O(w)\,dz.
\end{equation}
Then
\begin{equation}
  \begin{split}
	Q^+x^i&=\psi^i,\qquad
	Q^+x^{\bar\imath}=0,\\
	Q^+\psi^i&=0,\qquad
	Q^+\psi^{\bar\imath}=\partial x^{\bar\imath},\qquad
	Q^+\tilde\psi^i=0,\qquad
	Q^+\tilde\psi^{\bar\imath}=0,
\end{split}
\end{equation}
with the obvious generalizations to $Q^-$ and $\widetilde Q^\pm$. Then
\begin{equation}
  \begin{split}
    Q^+A(z,\bar z) &= \sum_j\frac{\partial
    f(x^i,x^{\bar\imath})}{\partial x^j} \psi^j\psi^{i_1}\psi^{i_2}\ldots \psi^{i_p}\tilde\psi^{\bar\imath_1}\tilde\psi^{\bar\imath_1}\ldots \tilde\psi^{\bar\imath_q}\\
    \widetilde Q^-A(z,\bar z) &= \sum_{\bar\jmath}\frac{\partial
                  f(x^i,x^{\bar\imath})}{\partial x^{\bar\jmath}}
                  \tilde\psi^{\bar\jmath}\psi^{i_1}\psi^{i_2}\ldots \psi^{i_p}\tilde\psi^{\bar\imath_1}\tilde\psi^{\bar\imath_1}\ldots \tilde\psi^{\bar\imath_q}.
  \end{split}  \label{eq:Qs}
\end{equation}
 
So the obvious similarity between (\ref{eq:ds}) and (\ref{eq:Qs})
shows we have a correspondence:
\begin{equation}
  \partiald = Q^+,\quad\bar\partiald = \widetilde Q^-,
\end{equation}
and the (c,a) fields thus correspond to $(p,q)$-forms. Given that
$k^{j,\bar j}$ was defined using (c,c) fields, we have
\begin{equation}
  h^{p,q} = k^{p,d-q}. \label{eq:sameHodge}
\end{equation}

Everything we have done today was done in a single flat coordinate
patch for torus. One can try to extend this to general Riemannian
manifolds but this is where all the difficulty of the geometry of
string theory comes into play. This involves nonlinear
$\sigma$-models, gauged linear $\sigma$-models, etc., which is well
beyond what we can cover here.

That said, we might generally hope that, given an $N=(2,2)$
superconformal field theory with the correct charge quantization
condition we might find some complex manifold $X$ whose Hodge numbers
agree with (\ref{eq:sameHodge}). In particular, this would imply that
\begin{equation}
  h^{d,0}=1,
\end{equation}
which is the \CY\ condition. Thus we expect a correspondence between
$N=(2,2)$ theories and \CY\ manifolds. But, then again, we have no
right to expect there to be a geometric interpretation of a particular
superconformal field theory and so this correspondence is far from
perfect.

This identification between differential forms of type $(p,q)$ with the
(c,a) fields is known as the A-model and naturally yields de Rham
cohomology. There is also a B-model story which goes via the (c,c)
fields. They are related by mirror symmetry, which, of course, is a
long story in itself!  

\subsection{Topological Field Theory}

The chiral primary fields have a nice ring structure thanks to
operator products and this contains a good deal of information about
the $N=(2,2)$ theory. In fact, it is surprising how much interesting
information can be extracted purely from just these chiral
fields. We can modify an $N=(2,2)$ theory in such a way as to
leave only this (c,c) information so we can hone in on this part of
the bigger $N=(2,2)$ picture. This leads to ``topological field
theory''. Much of the modern analysis of string theory in the context
of geometry is presented as a topological field theory so we will
present it here, albeit very briefly.

The last term in (\ref{eq:N=2OPE}) suggests a new field
\begin{equation}
  F(z)=T(z)-\ff12\partial J(z),
\end{equation}
might be an interesting object. In terms of modes we write
$F_n=L_n-\frac 12 (n+1)J_n$.

Let us use this to eliminate $T(z)$ in favor of $F(z)$ in the $N=2$
algebra. Furthermore, let's rename $G^\pm(z)$ to $Q(z)$ and $G(z)$
respectively. We then obtain
\begin{equation}
  \begin{split}
	F(z)F(w)&=\frac{2 F(w)}{(z-w)^2}+\frac{\partial F(w)}{z-w}+\dots\\
	F(z)Q(w)&=\frac{Q(w)}{(z-w)^2}+\frac{\partial Q(w)}{z-w}+\dots\\
	F(z)G(w)&=\frac{2G(w)}{(z-w)^2}+\frac{\partial G(w)}{z-w}+\dots\\
	F(z)J(w)&=\frac{d}{(z-w)^3}+\frac{J(w)}{(z-w)^2}+\frac{\partial
                  J(w)}{z-w}+\dots\\
	Q(z)G(w)&=\frac{2d}{(z-w)^3}+\frac{2J(w)}{(z-w)^2}+\frac{2F(w)}{z-w}+\dots  \\
	J(z)Q(w)&=\frac{Q(w)}{z-w}+\dots \\
	J(z)G(w)&=-\frac{G(w)}{z-w}+\dots \\
	J(z)J(w)&=\frac{d}{(z-w)^2}+\dots 
\end{split}
\end{equation}

Now let's try to reinterpret this as a conformal field theory with
stress tensor $F(z)$. Note that by doing this, we really are
considering an inequivalent conformal field theory. The first equation
tells us that the new central charge is 0. But this is supposed to be
a measure of the complexity of the conformal field theory, so clearly
our new theory is very simple!

Indeed, if $c=0$ then a positive inner product on the Hilbert space
forces the condition that $F_n\ket\psi=0$ for any $n$
\cite{Gins:lect}. So any field is completely invariant under
reparametrization. The fact that $F_0\ket\psi=0$ is precisely that
$h-\ff12j=0$ and so the only fields in our new theory are chiral
primary fields from the old theory. The vast majority of fields from
the original $N=2$ theory have been lost.

When we simply ``twist'' the $N=2$ algebra to the topological version
above, the adjoint of $Q(z)$ is $G(z)$. One can also choose to alter
the inner product structure such that $Q(z)$ becomes self-adjoint. The
positive definiteness of the inner product is now lost unless we restrict
attention to $Q$-cohomology. In this case the setup looks similar to
BRST quantization and, because of this, people often refer to the
R-charge as ``ghost number'' instead.

The condition $F_{-1}\ket\psi=0$ is the condition that translations in
$z$ are invariant. This means that any correlation function will not
depend on the location of vertex operators (at least so long as they
don't collide or hit singularities in the Riemann surface). This is
why the theory is called ``topological''.

The chiral ring structure of the original $N=2$ theory is preserved in
this topological field theory and this gives a context in which it is
much easier to analyze. Sadly, we have no time to pursue this further.

\section*{Acknowledgments}

I would like to thank Owen Gwilliam and the organizers of the
Physical Mathematics of Quantum Field Theory Summer School. I am also
very grateful to Muldrow Etheredge for taking notes and writing up the
initial form of these notes.

\end{document}